\documentclass[letterpaper, 10 pt, conference]{ieeeconf}  

\usepackage{graphicx}
\graphicspath{{images/}}
\usepackage{amsmath}
\usepackage{subcaption}
\usepackage[thinlines]{easytable}
\usepackage{cite}

\IEEEoverridecommandlockouts                              

\title{\LARGE \bf
A Sketch-Based System for Human-Guided Constrained\\Object Manipulation
}

\author{Sina Masnadi$^{1}$, Joseph J. LaViola Jr.$^{1}$,\\Xiaofan Zhu$^{2}$, Karthik Desingh$^{2}$ and Odest Chadwicke Jenkins$^{2}$
\thanks{$^{1}$Department of Electrical Engineering \& Computer Science, University of Central Florida,
        {\tt\small \{masnadi, jjl\}@cs.ucf.edu}}
\thanks{$^{2}$Department of Electrical Engineering \& Computer Science, Robotics Institute, University of Michigan, Ann Arbor
        {\tt\small \{zhuxf,kdesingh,ocj\}@umich.edu}}
}

\begin{document}

\maketitle
\thispagestyle{empty}
\pagestyle{empty}

\begin{abstract}

In this paper, we present an easy to use sketch-based interface to extract geometries and generate affordance files from 3D point clouds for robot-object interaction tasks. Using our system, even novice users can perform robot task planning by employing such sketch tools. Our focus in this paper is employing human-in-the-loop approach to assist in the generation of more accurate affordance templates and guidance of robot through the task execution process. Since we do not employ any unsupervised learning to generate affordance templates, our system performs much faster and is more versatile for template generation. Our system is based on the extraction of geometries for generalized cylindrical and cuboid shapes, after extracting the geometries, affordances are generated for objects by applying simple sketches. We evaluated our technique by asking users to define affordances by employing sketches on the 3D scenes of a door handle and a drawer handle and used the resulting extracted affordance template files to perform the tasks of turning a door handle and opening a drawer by the robot.

\end{abstract}

\section{Introduction}

Humans have the ability to explore objects around them and learn about ways they can interact with the objects. Objects may be acted upon in various ways. For instance, a human knows a cup can be picked up by its handle or be used as a liquid container. As we grow, we learn about these properties but how we learn these properties is still one of the mysteries of psychology. In robotic interaction systems, providing these interaction data and structuring them can bring a huge progress to object manipulation tasks.

Abstracting interactions with objects and manipulating them has been a problem in psychology. Affordances concept appeared in the early sixties to as an alternative perspective for visual perception. This theory has since expanded to object manipulation. The theory of affordances was introduced by J. J. Gibson for the first time \cite{Gibson1979GibsonPrint} to describe the interactions between organisms and their environments. 

An \textit{affordance} is described as the actions an object can afford. However, it does not necessarily define the object itself. For example, when we say an object has sitting affordance, it does not necessarily mean it's a chair; it can be a chair, a sofa, a bench, or even the floor \cite{Hart15}. Affordances help us create concepts for the object manipulation tasks and create high-level partitions for interactions. Hart et al. provided the affordance template ROS package for robot task programming which is an attempt to standardize robotic interaction programming \cite{Hart15}.

Most human-robotic interaction systems are based on complicated 3D interaction and programming systems which require expert knowledge to program and operate them. This can pose a problem for novice users who need to use a robot to perform object manipulation tasks. Our system provides numerous advantages over the conventional systems to interact with robots by utilizing sketch-based techniques to provide a straightforward user interface for the user to interact with a robot. Also, we leverage human-in-the-loop to provide affordance-based interactions in our approach to manipulate objects.

Sketching is a low overhead means of rapid communication. The main idea behind our approach is utilizing human-in-the-loop, to define object affordances with simple sketches and improve the autonomous tasks precision. Human-in-the-loop technique will provide us the ability to aid the robot and correct its faulty behavior when it demands direction or faces unresolvable problems. In contrast to the fully autonomous systems which can be a black box for the user, human-in-the-loop can boost the performance of the robot by presenting the right data at the right time to the user and increase the reliability of the performed tasks.


The geometry of objects is an important factor which can aid the robot's sensing and perception. Model based generative approaches such as \cite{karthik, narayanan2016discriminatively, mahler2017dex, sui2017sum} used mesh models that are synthetically designed or scanned using 3D scanners. Perceptual tasks in object manipulation systems rely on pose estimation and the mesh reconstruction. Using human sketches on top of the scanned 3D scene for extracting the 3D mesh of the objects is one of the approaches which can be useful in cluttered scenes. Maghoumi et al. introduced a sketch-based system to extract 3D geometry from point clouds in cluttered scenes \cite{Maghoumi2018GemSketchClouds}. We also have used the same approach to extract the geometries as the prerequisites of our work.

%
%

\section{Related Work}
\label{sec:related}

Object manipulation tasks can be divided into three main steps: perceiving the environment, structuring the actions, and path planning and performance. Perceiving the environment is basically using sensors to scan the environment and create a map of the world that robot can interact with. Structuring the actions is defining the sequence of actions for the robot to perform for a specific task. Path planning and performance is deciding which moves should be taken to achieve the actions and ensuring those actions has completed successfully. Affordances concept can create a structure for robot actions. A robot can use the perceived environment data to explore the actions that can be performed on the objects and the affordances those objects can have.

Affordances have a similar definition in major fields such as psychology, neuroscience and robotics \cite{jamone2018affordances}. Gibson's theory of affordances divides environmental objects into two main categories of attached and detached \cite{Gibson1979GibsonPrint} objects. Attached objects are partially or wholly surrounded by the medium which cannot be displaced without becoming detached. Detached objects are objects that can be displaced and afford carrying. Detached objects must be relative to the size of the subject to be moved or handled. The similarity of Gibson's theory of affordances with what robots need to know to interact with objects was initially discussed by Artkin \cite{brooks1990elephants} and Sahin et al. \cite{csahin2007afford}. During the past twenty years many roboticists have successfully used affordance related ideas to implement intelligent robotic systems \cite{jamone2018affordances}.

Using affordances in robotics is based on two major approaches, providing the robot with the pre-processed affordances or training the robot to learn affordances of the objects by interacting with them. We are going to focus on the first approach in our system, which uses human inputs to fabricate the affordances of an object. The second approach uses artificial intelligence techniques to create a cognitive model for robots to understand the objects' affordances.

In robotics, the focus is mainly on affordances of detached objects. Stoytchev \cite{Stoytchev2005Toward} examined a developmental approach to learn the binding affordances of objects by a robot. In their work different objects affordances were represented in terms of a set of exploratory behaviors and their observed effects. Ugur et al. \cite{ugur2011goal} showed that an anthropomorphic robot equipped with a range camera can learn object affordances through self-interaction and self-observation (unsupervised learning) and use these affordances for planning. Another significant work has been done by Sun et al. \cite{sun2010learning} which is a probabilistic graphical model that they call affordance category model. They use this model to introduce learning and training strategies for robots.
However these approaches could relate the behaviors to different objects, the training time is significant and they cannot be related to more complex interactions.

Providing the robot with pre-processed affordances has been discussed in many robotic researches. Hart et al. \cite{Hart15} provided a template for affordances which can be used with ROS. The affordance template is an environment to quickly program, adjust, and execute robot applications in the ROS RViz environment. Marion et al. provided a user interface for robot operation with shared autonomy \cite{marion2017director}, to interact with robot and fitting affordances on the 3D scene. Our approach differs from theirs' in that we create the affordances by simple sketches on the objects and in geometry extraction, which their system uses convex hull to create meshes, but our system is based on generalized cylindrical shapes and cuboids.

%
%
\section{Overview of The System}
\begin{figure}[h]
    \centering
    \includegraphics[width=\linewidth]{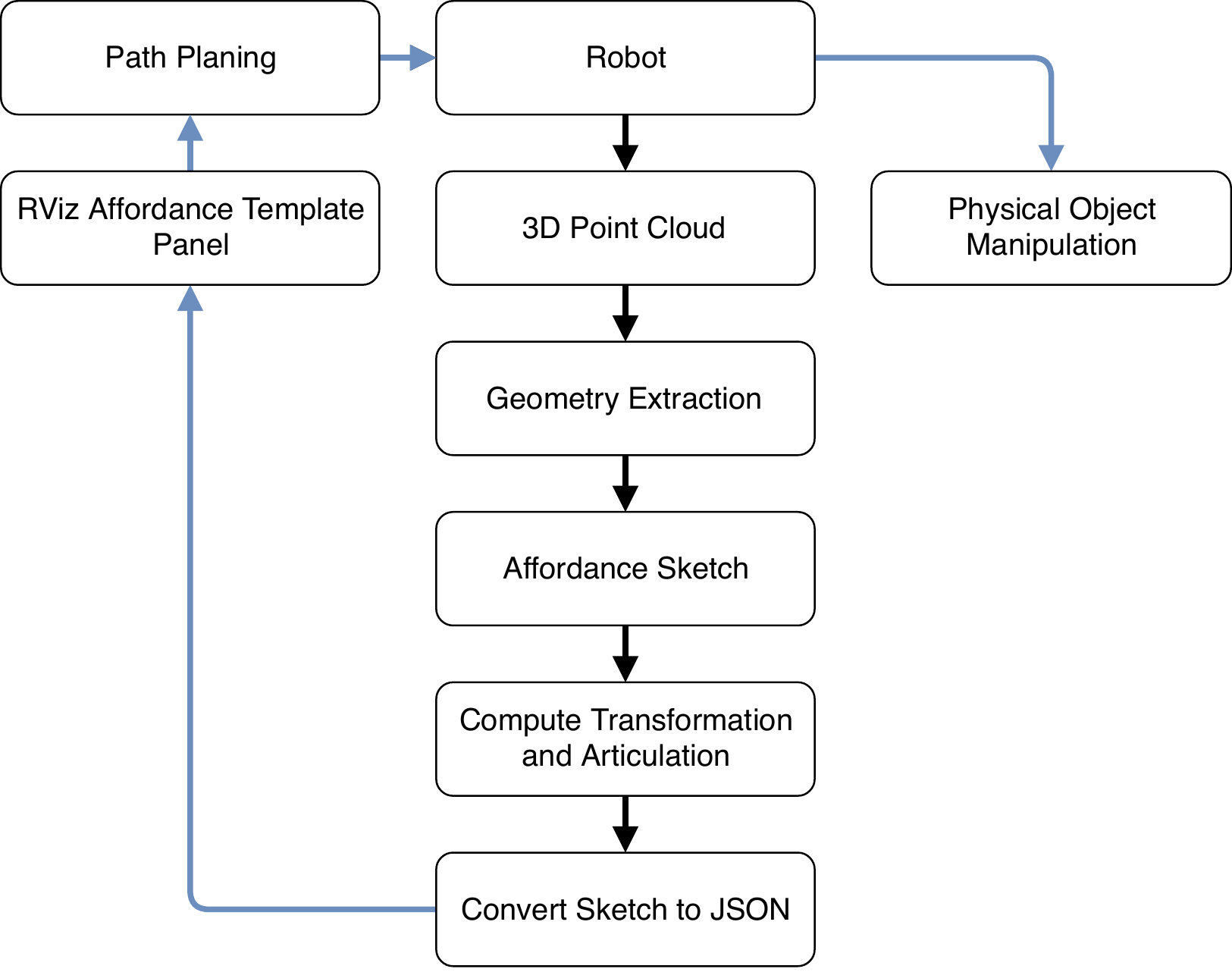}
    \caption{Sketch-based affordance procedure}
    \label{fig:procedure}
\end{figure}

In this section, we are going to discuss a high level overview of how the system works and its components are connected. The main steps of the system are: Capturing the scene using robot's depth camera, Geometry extraction, Affordance extraction, Affordance template generation, RViz affordance template placement and having the robot to perform the task. Figure \ref{fig:procedure} shows the procedure of performing a task by the robot and our sketch-based affordance extraction system. 

In the next sections we are going to discuss how we can extract the geometry of the objects in the scene, how can the user sketch the affordances, the process of converting the user's provided data to affordance templates, and how these generated data are used to perform a task on the robot. Finally we are going to discuss the results for the sample task performance using the robot.

\section{Geometry Extraction}\label{sec:geometryextraction}

Maghoumi et al. presented GemSketch, an interactive system for extracting the geometries of generalized cylinders and cuboids from single-view or multiple-view point clouds \cite{Maghoumi2018GemSketchClouds}. This system enables the user to extract geometry of the objects by simply sketching their silhouette. We extended their code to fit to affordance extraction purpose. Our extension can now be used to assign affordances to objects and group multiple extracted objects into one single object.
To extract common objects such as door handles or drawer handles, we defined a new optimized geometry extraction tool based on the existing. This tool uses the 3D surface of the object's ground (like the door surface or a drawer surface) and fits a plane onto that surface and uses it as a base for the profile plane. The user will identify the surface by drawing a curved line on it and then a 3D plane is fitted to that line's points using RANSAC. This plane helps GemSketch to provide a pose estimation for objects that are perpendicular to other surfaces. Then the geometry extraction is continued similar to GemSketch approach.

\begin{figure} [h]
    \centering
    \includegraphics[width=\linewidth]{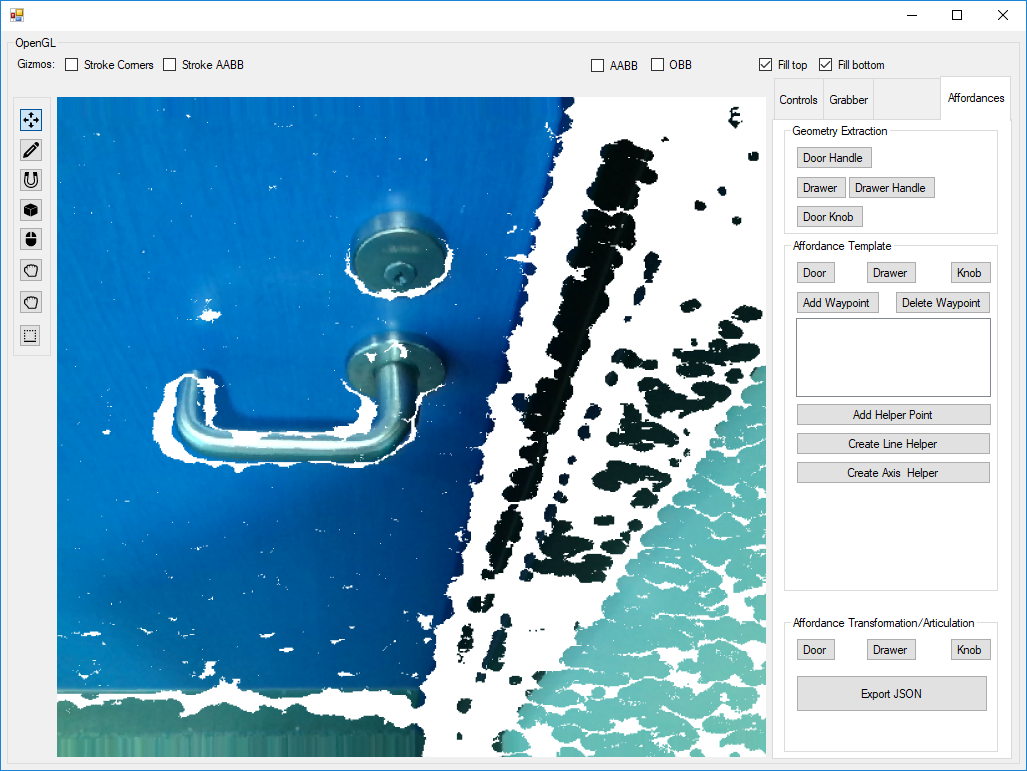}
    \caption{Geometry and affordance extraction user interface}
    \label{fig:my_label}
\end{figure}

\section{Affordance Extraction}\label{sec:affordanceextraction}
\subsection{Overview}
Our system provides tools to process and import the output of the robot's depth camera which is a 3D point cloud and its accompanying 2D image. In this work, we extended our sketch based affordance tools \cite{masnadi2019sketching} to provide the user with the necessary tools for affordance extraction. User has the option to draw on the 3D point cloud or the 2D image which is projected to the 3D point cloud. User can see a list of the extracted objects and can define affordances on the selected object. We will discuss the process of sketch-based affordance extraction in the following section.

\subsection{Sketch Details}
Sketching of the affordances is based on affordance waypoints. An \textit{affordance waypoint} defines the position, orientation, status of the robot's gripper (open/closed) in the 3D environment, and two other features, type and axis. Affordance waypoint's \textit{type} shows if it is a \textit{direct} or an \textit{assistive} affordance waypoint.
There are two ways to create affordance waypoints, direct and assisted.

\begin{itemize}
\item \textit{Direct Affordance Waypoints}\\Users can add direct affordance waypoints by directly drawing a circle on the scene's point cloud or image. We use the circle to project a ray on the point cloud and add the affordance waypoint where the ray hits the point cloud.
\item \textit{Assisted Affordance Waypoints}\\Assisted affordance waypoints are created using \textit{helpers}. A \textit{helper} is a set of points, lines, and axes which can help the user to move the affordance waypoints or create them based on constraints.
\end{itemize}

Our system has the ability to add new helpers. Currently, we have provided three helpers to the user. These helpers are perpendicular helper, parallel line helper, and rotation axis helper. Given these helpers user can achieve a combination of affordance waypoints to create a certain affordance. Table \ref{tab:helpers} illustrates the different types of helpers.

Creating helpers starts with adding a helper point to the scene. Helper points are points which can be used to create perpendicular helpers, parallel line helpers, and rotation axis helpers. A perpendicular line helper will help the user to add waypoints perpendicular to a surface or a line. A parallel line helper will help the user to add waypoints along its direction. A rotation axis helper will help the user to rotate an affordance waypoint around its axis to create a new affordance waypoint.

Our system currently supports sketching the interactions which consists of opening and closing the robot gripper, moving the end effector in a straight line, and rotating the end effector around a specified center of rotation. These interactions can be combined for generating more complex interactions such as using door handle to open doors and opening and closing drawers.

\begin{table}[h]
    \centering
    \begin{TAB}(c,0.1cm,2cm)[5pt]{|c|c|}{|c|c|c|c|}
        Helper Type & Illustration \\
        \begin{minipage}{.45\linewidth}
        \textbf{3D perpendicular helper}
        Users can use perpendicular line helper to quickly create an affordance waypoint which is on a perpendicular line to a surface.
        \end{minipage}
         & 
        \begin{minipage}{.3\linewidth}
            \includegraphics[width=\linewidth]{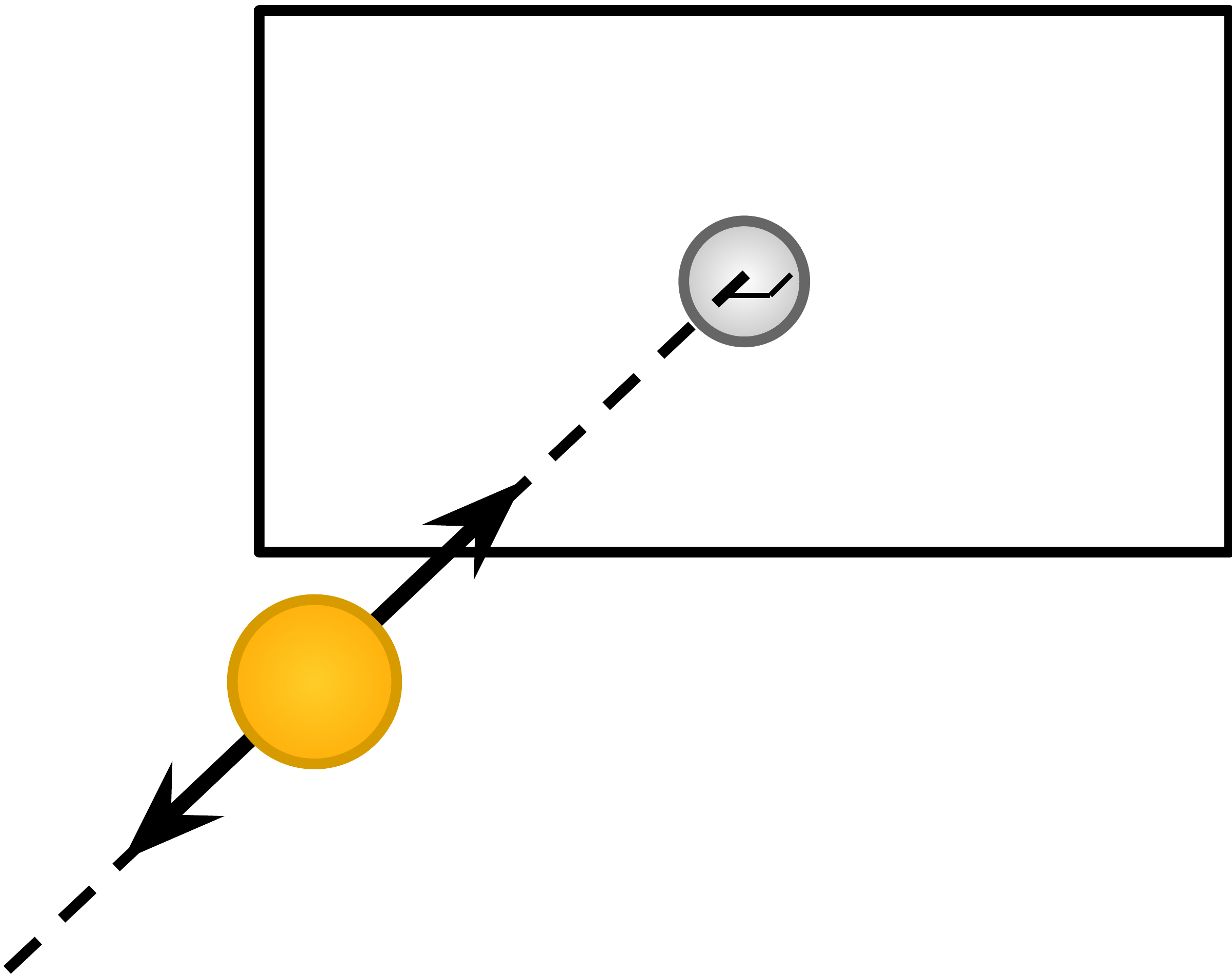}
        \end{minipage}\\
        
        \begin{minipage}{.45\linewidth}
        \textbf{3D parallel line helper}
        Users can use it to create a line $l$ from two other helper points or affordance waypoints to create and move an affordance waypoint on a line parallel to $l$. 
        \end{minipage}
        &  
        \begin{minipage}{.3\linewidth}
            \includegraphics[width=\linewidth]{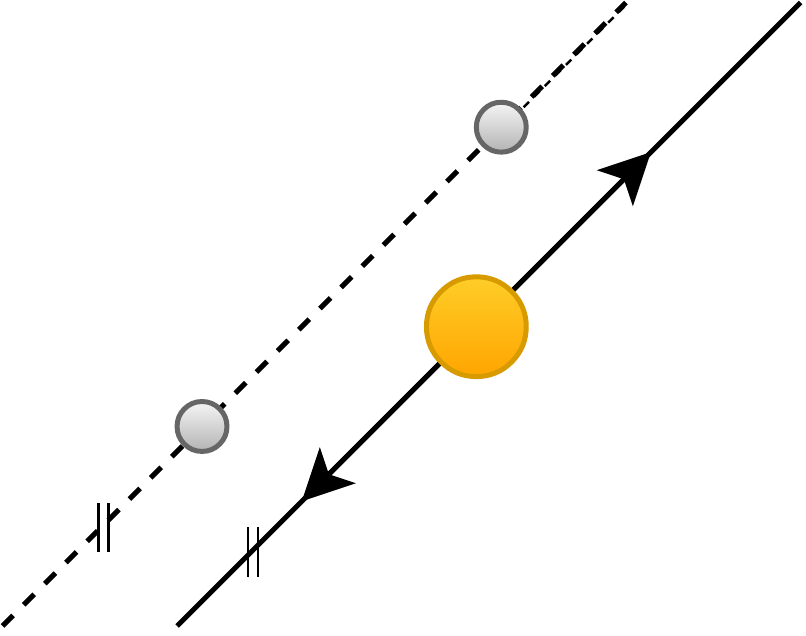}
        \end{minipage}\\
        \begin{minipage}{.45\linewidth}
        \textbf{3D rotation axis helper}
        This helper can be used to create an axis line $l$ and create a new affordance waypoint (or select an existing one) and rotate it around $l$. 
        \end{minipage}&  
        \begin{minipage}{.3\linewidth}
            \includegraphics[width=\linewidth]{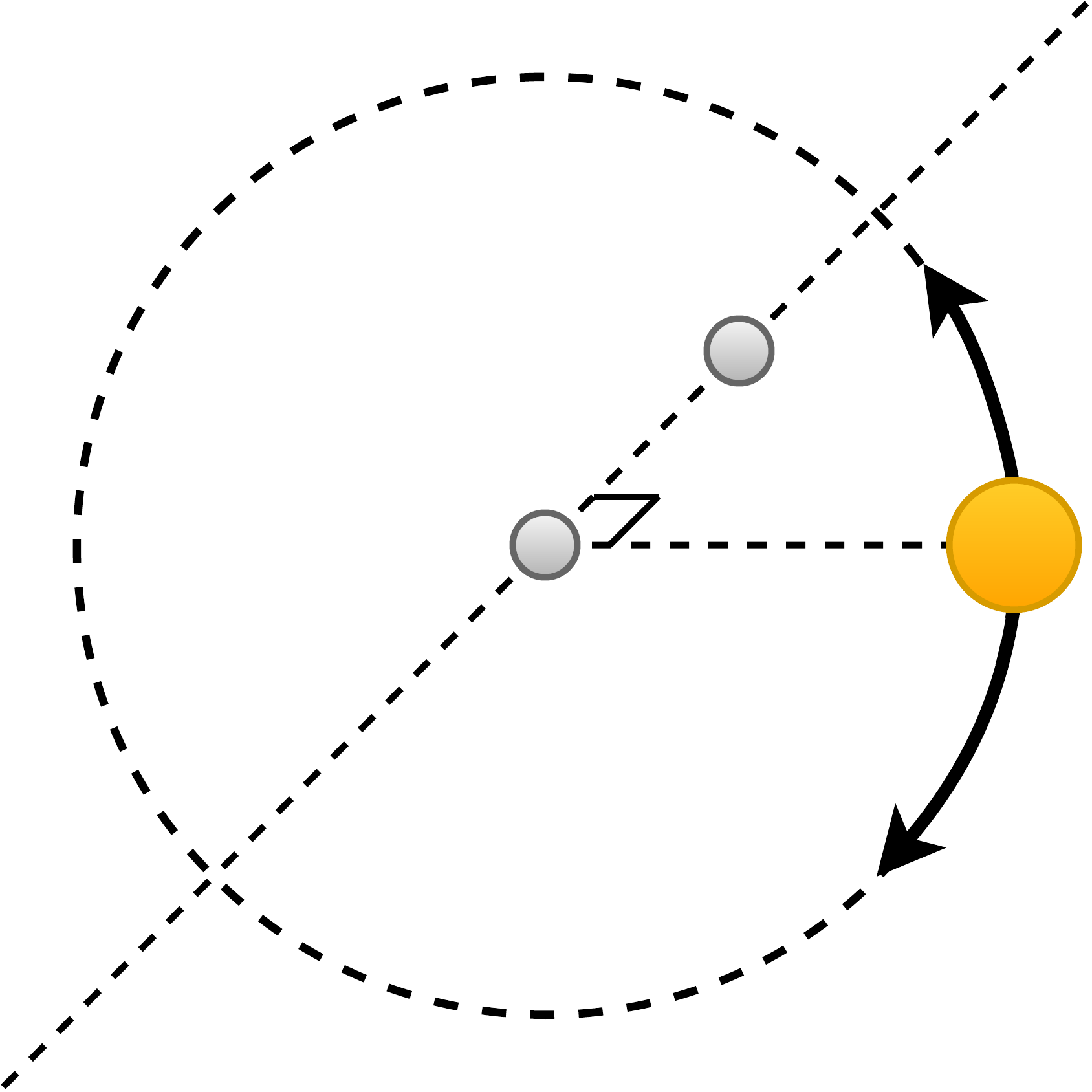}
        \end{minipage}\\
    \end{TAB}
    \caption{Illustration of different types of helpers. Gray points are helper points and orange points are the affordance waypoints created using the helpers.}
    \label{tab:helpers}
\end{table}

After the user created the list of affordance waypoints, they should add a point to define the origin of the affordance which will also be used as the origin of object's geometry. Origin of affordance waypoints and the geometry will be translated to the new origin and the data from the list is converted to a JSON file that will be used to create the affordance template for the object. We will discuss the affordance templates in the next section.

\section{Affordance Templates}\label{sec:executing}

\subsection{Affordance Template Introduction}

Hart et al. introduced the affordance template which is a JSON representation of an object's affordances. An affordance template consists of two main parts, $G_{obj}$ being the geometry information of an object and $F_{JSON}$ being the poses of the geometry and all related waypoints. $F_{JSON}$ is based on object poses, which means the procedure that the robot carries to perform a task is divided into poses and each pose is defined by the following: status of robot gripper being open or closed, gripper's position and orientation which are based on the pose of the object's geometry.

Based on the existing affordance template package \cite{Hart15}, the $G_{obj}$ should be stored in a mesh file and $F_{JSON}$ = $ \{F_{geo}, F_{wp} \}$ written in JSON. $F_{geo}$ is the geometry pose with respect to the pose of the robot, which affects the display of our geometry in RViz when loading the template. $F_{wp}$ contains the poses of all desired waypoints, which would be followed by the robot gripper step by step. By adding these two files into the package, we could create a new affordance template. Executing the new template in RViz affordance template panel would enable the robot to move accordingly as the user sketches.

In our work, $G_{obj}$ is given by our sketching system. To generate $F_{JSON}$ requires pose of geometry and poses of waypoints which are compiled by taking the user sketches as inputs. The required poses of waypoints include stand-alone waypoints $F_{sta}$ that are translating the gripper position but keeping its orientation and rotational waypoints $F_{rot}$ that is changing gripper position and orientation accordingly at the same time.

\subsection{JSON generation}

The affordance template is a computational construct that can be used in a graphical environment to utilize human inputs to guide a robot to its goals. it is an attempt to standardize programming robotic manipulation. However, creating JSON files for a template can be a time consuming task which requires users to carefully fabricate the template, such as typing numbers and names into files. This could be especially complicated and risky when the manipulation task is complex.  Our approach is addressing this problem by introducing a compiler for the list of affordance waypoints that not only write all the stand-alone waypoints, but also interpolate intermediate waypoints \cite{Shoemake1985} that are required for the  robot to perform the tasks, but are not sketched in the list. 


As mentioned above, $F_{wp}$ consists of $F_{sta}$ and $F_{rot}$. All the stand-alone waypoints $F_{sta}$ should be completely defined by user sketches, while the rotational waypoints $F_{rot}$ could be defined by only requiring a start pose and end pose, instead of defining them one by one. An interpolation method is implemented here to fill out the intermediate poses between start and end pose, which finishes $F_{rot}$ in a JSON file.

The interpolation of rotational waypoints uses spherical linear interpolation (slerp) in Kuffner J.J.'s work \cite{Kuffner2004}, which takes the interpolation portion, start quaternion and the end quaternion as input and returns the quaternion at a given interpolation portion. 

Slerp could interpolate quaternions using an incremental portion which is the reciprocal to the number of waypoints in the manipulation. Users can define the number of waypoints desired in the whole manipulation trajectory. When generating the JSON file, our method will recursively generate the waypoint quaternions along the manipulation trajectory desired. Finally, by applying transformation of interpolated quaternions on the start pose position, $F_{wp}$ would have the information of every rotational waypoint. Similar sessions to the above session could be performed on other manipulations, generating the JSON file of poses $F_{wp}$ with all the rotational waypoint poses $F_{rot}$ according to user sketches.

\begin{figure*}
  \centering
    \begin{subfigure}[b]{0.23\linewidth}
        \includegraphics[width=\linewidth]{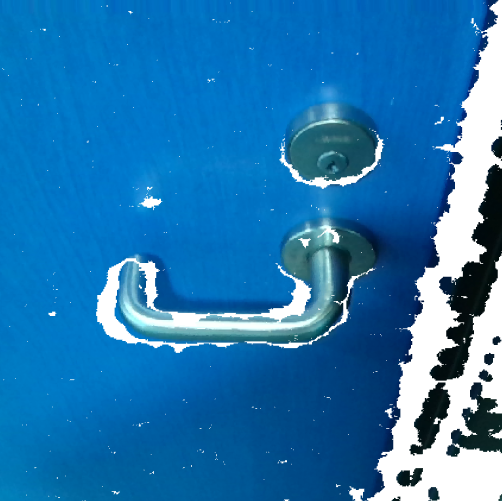}
        \caption{Point cloud view of the door and its handle}
        \label{fig:handle1}
    \end{subfigure}%
    \hspace{0.02\linewidth}
    \begin{subfigure}[b]{0.23\linewidth}
        \includegraphics[width=\linewidth]{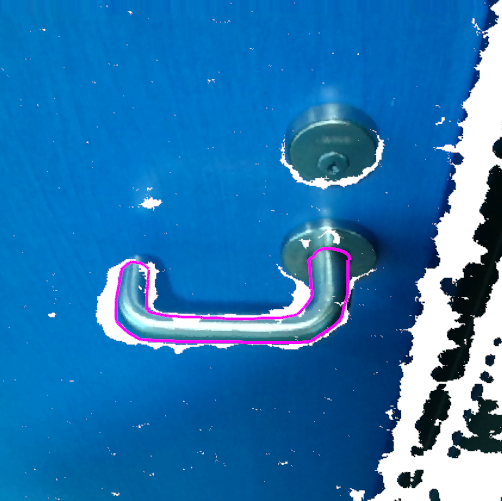}
        \caption{Traced silhouette of the door handle}
        \label{fig:handle2}
    \end{subfigure}%
    \hspace{0.02\linewidth}
    \begin{subfigure}[b]{0.23\linewidth}
        \includegraphics[width=\linewidth]{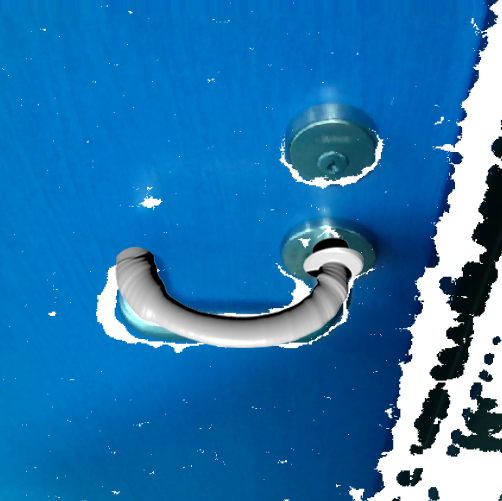}
        \caption{Extracted door handle geometry from user's sketch}
        \label{fig:handle3}
    \end{subfigure}%
    \hspace{0.02\linewidth}
    \begin{subfigure}[b]{0.23\linewidth}
        \includegraphics[width=\linewidth]{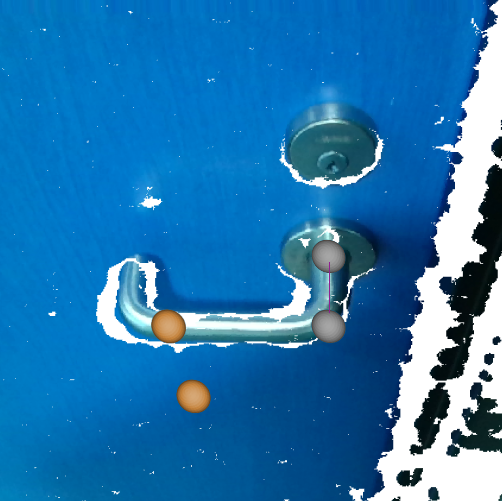}
        \caption{User provided waypoints and helper points for the door handle}
        \label{fig:handle4}
    \end{subfigure}%
    \caption{Tracing the outlines of the door handle to extract its 3D mesh. The user then adds affordance waypoints to the scene.}
    \label{fig:handlesilhouette}
    \vspace{-0.5cm}
  
\end{figure*}

With $F_{sta}$ generated from user sketches and interpolation of $F_{rot}$, $F_{wp}$ is ready and serves as our JSON template. The JSON file and the extracted geometry in section \ref{sec:geometryextraction} forms the desired affordance template for robot to perform tasks.

\subsection{Execution}

After adding the new template to the affordance template package, we are ready to load and execute the template in RViz. As shown on the right side of Fig \ref{fig:affordance1}, the geometry and waypoints are loaded corresponding to the user defined $F_{geo}$. 

With the correct template loaded, which is shown on the left side of Fig \ref{fig:affordance1}, the geometry needs to be manually registered to the perception view of robot's camera. The motion planner in MoveIt! will then start planning on all the waypoints $F_{wp}$ to see if a motion plan is available. If true, by clicking on "execute" button, we can tele-operate the robot to move either waypoint by waypoint or by several waypoints.

\section{Experimental Results}
\label{sec:results}
We used our system to extract the geometry and affordance of a door handle and a drawer handle. The affordance of a handle shows how it can afford being opened. The mesh of the door handle is considered a generalized cylinder and is extracted by drawing the silhouette of the door handle. Figures \ref{fig:handle2} and \ref{fig:handle3} display how user's sketches on the point cloud are converted to the mesh of the door handle. We extracted the geometries and the affordances ten times for each of the objects. 

We measured the accuracy of the performance using all-or-nothing approach \cite{wolin2008shortstraw}. The accuracy here is calculated by dividing the number
of correct robot task performance divided by the total number of
extracted affordances.

We are going to first describe how the affordances are extracted for the door handle and the drawer, then we will discuss how the task is performed on the robot, finally, the success and failure of the robot is analyzed.

\begin{figure}[h]

\end{figure}

\subsection{Sketching The Affordance}
\subsubsection{Door Handle}
The door handle affordance is defined by its rotational affordance that results in opening of the door. To rotate a door handle, the robot must first know where to grab, the second step is rotating the door handle around its axis of rotation. To rotate the door handle, the robot needs to know the axis of rotation.

The first step of extracting the door handle affordance is segmenting the door geometry to door surface and everything that is located on top of this surface. The door surface can be extracted by asking the user to draw a curved line on the door. Since the line consists of a set of points $points_d$, we can fit a plane to these point using RANSAC. We call the door surface plane $P_D$. The user will then select the area that robot should grasp in order to grab the door handle to rotate it and open the door. The selected area mean will be used as the grasping point. We refer this point as $p_g$ which is an \textit{affordance waypoint}. The next step is using a \textit{3D rotation axis helper} to create a new affordance waypoint for the door handle opening projection. Using $p_g$ and $P_D$, we calculate a plane $P_H$ parallel to $P_D$ which contains $p_g$, this plane is going to help us extract the axis of rotation for the door handle and the projection of the door handle when the robot is going to open the door. Then user selects the point that they think the axis of the door handle is located on and creates a \textit{helper point}, which we call $p_{ah}$. The projection of $p_{ah}$ on $P_D$ is called $p_{ad}$ is another helper point. Using $p_{ah}$ and $p_{ad}$, the 3D rotation axis helper is built and we call this rotation axis $L_r$.
\begin{figure}[h]
    \centering
    \includegraphics[width=\linewidth]{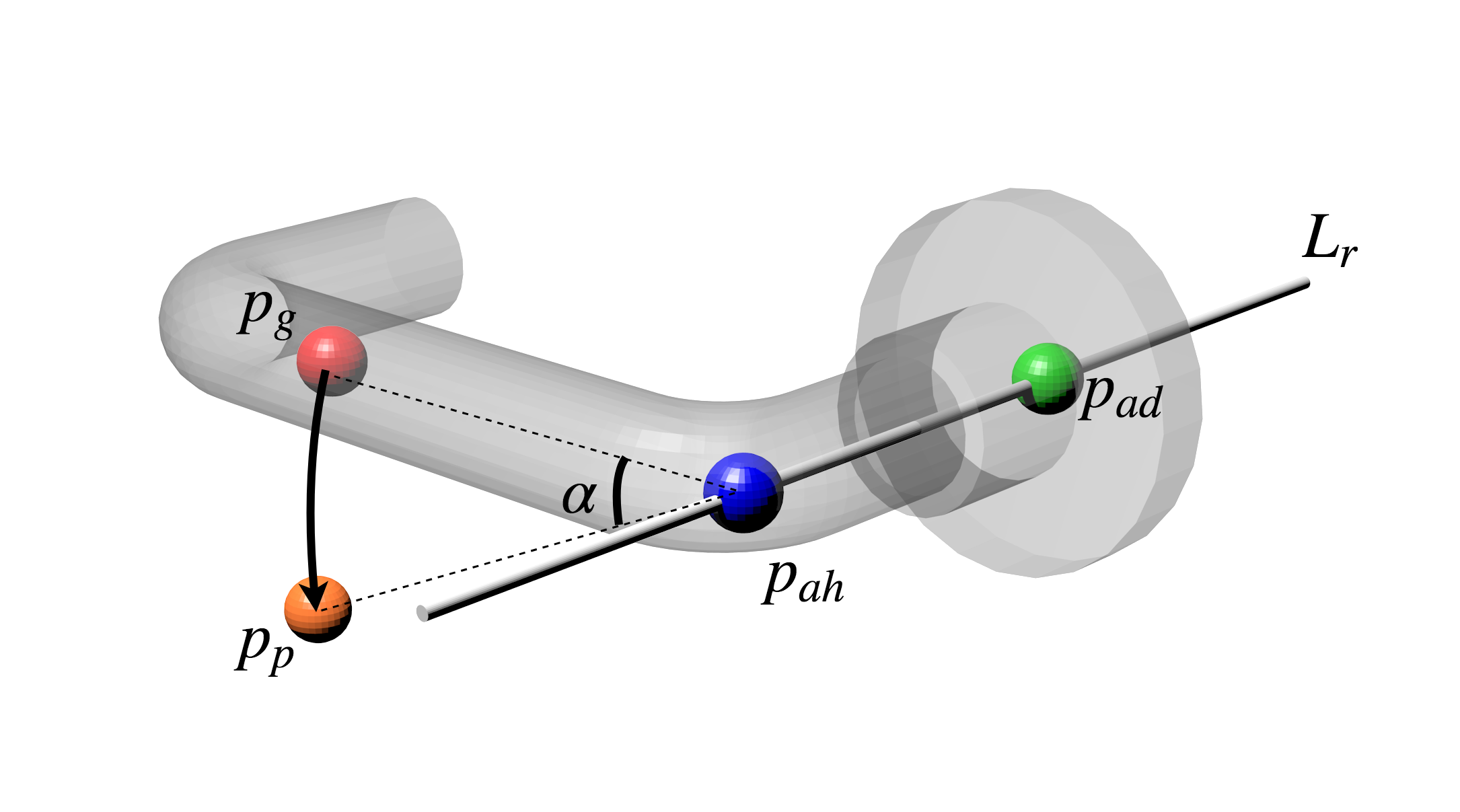}
    \caption{Door handle transformations and articulations.}
    \label{fig:doorhandleTA}
\end{figure}
Now that we have the door handle grasp location and its axis of rotation, the user is asked to draw the projection of the door handle and show the system where the door handle should be moved to, to retract the door latch. Using $L_r$ as a helper axis, by grabbing the grasp location, it can be rotated temporarily around $L_r$, when user reaches a point that they think is a good projection for the door handle, they release the grasp location and the system is going to save the projection point as $p_p$. The angle of rotation of the door handle ($\alpha$) will be used to process the robot's gripper for task performance. Figure \ref{fig:doorhandleTA} illustrates the data collected from the user.

\begin{figure*}
  \centering
    \begin{subfigure}[b]{0.3\linewidth}
        \includegraphics[width=\textwidth]{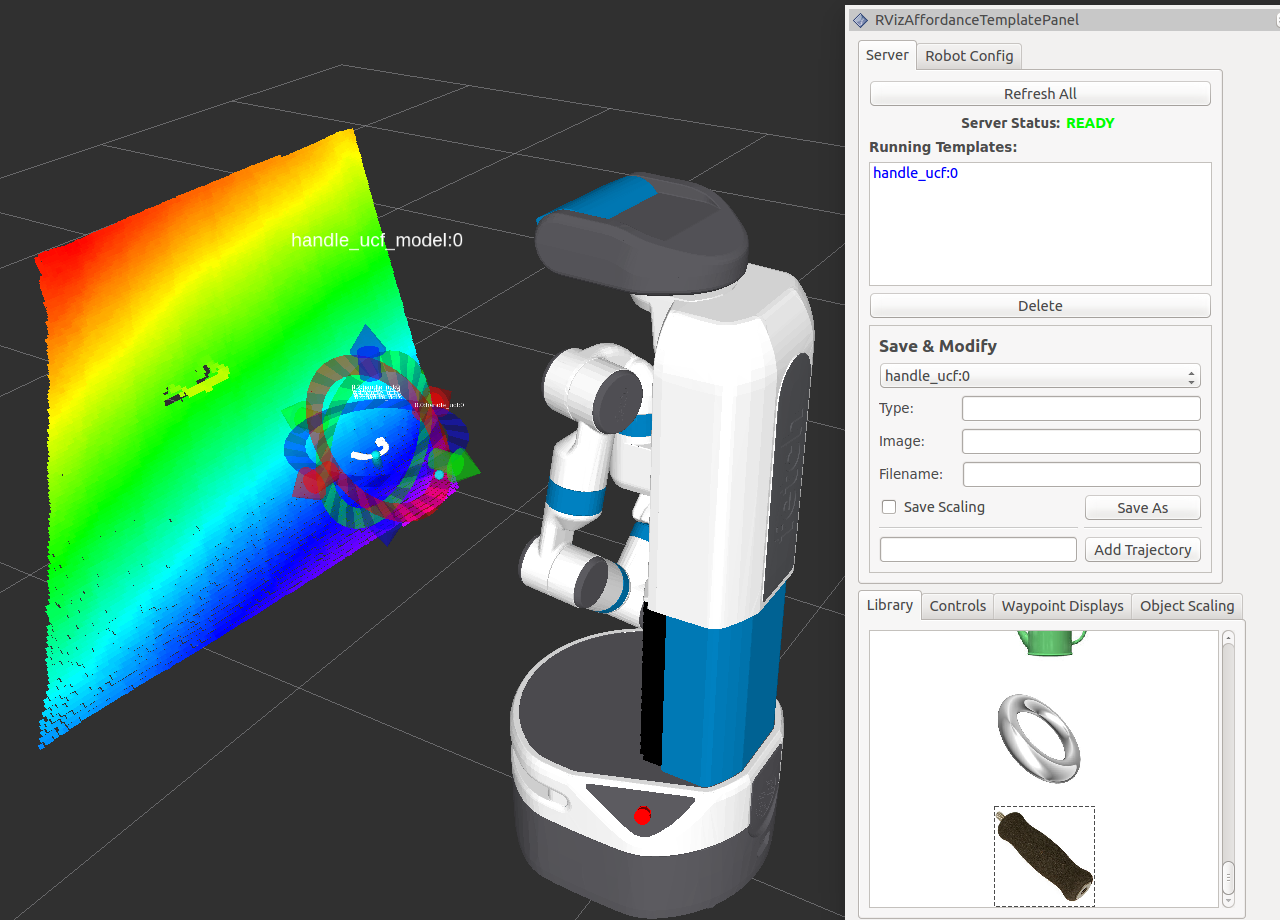}
        \caption{Loading template from panel}
        \label{fig:affordance1}
    \end{subfigure}%
    \hspace{0.05cm}
    \begin{subfigure}[b]{0.3\linewidth}
        \includegraphics[width=\textwidth]{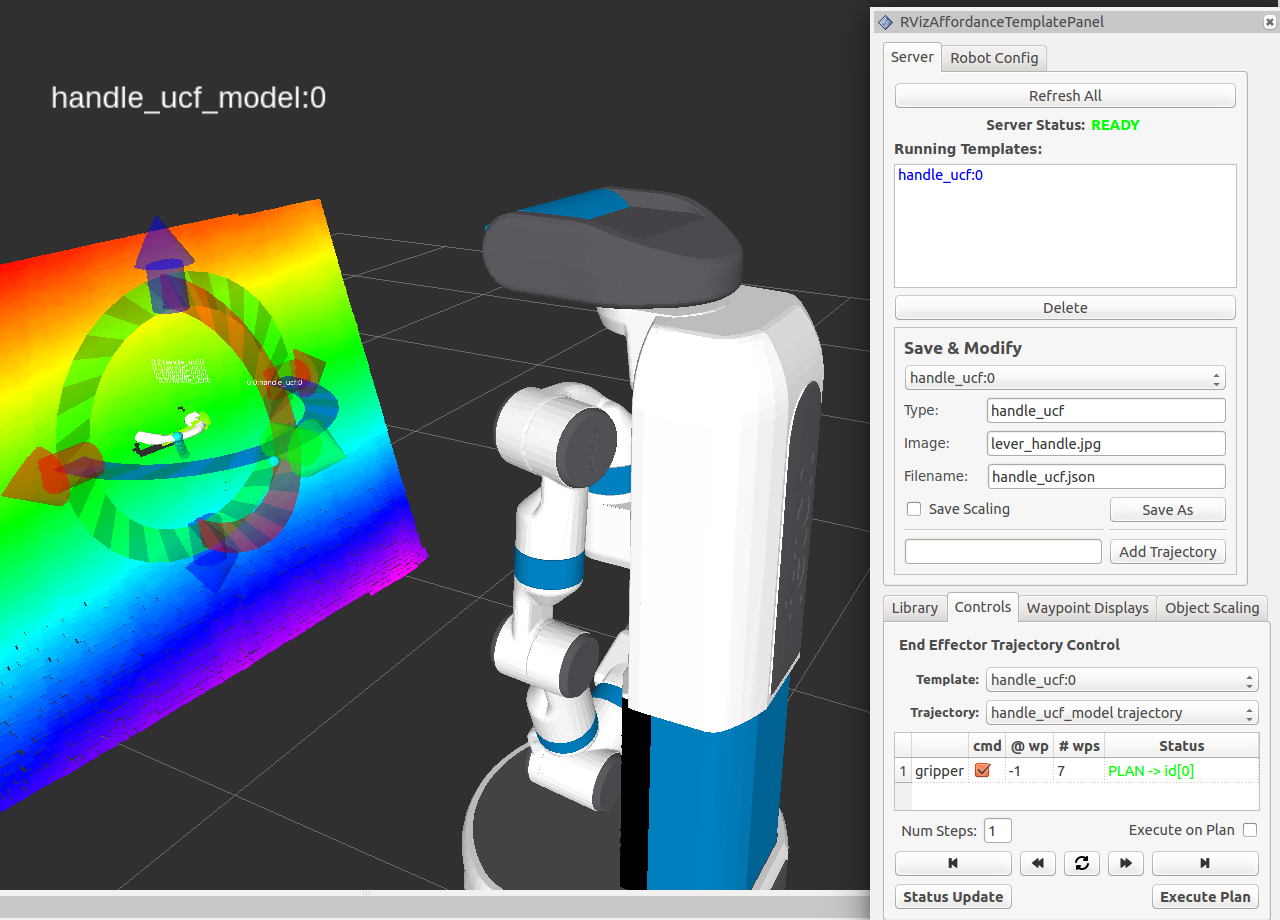}
        \caption{Manually fitting geometry}
        \label{fig:affordance2}
    \end{subfigure}%
    \hspace{0.05cm}
    \begin{subfigure}[b]{0.3\linewidth}
        \includegraphics[width=\textwidth]{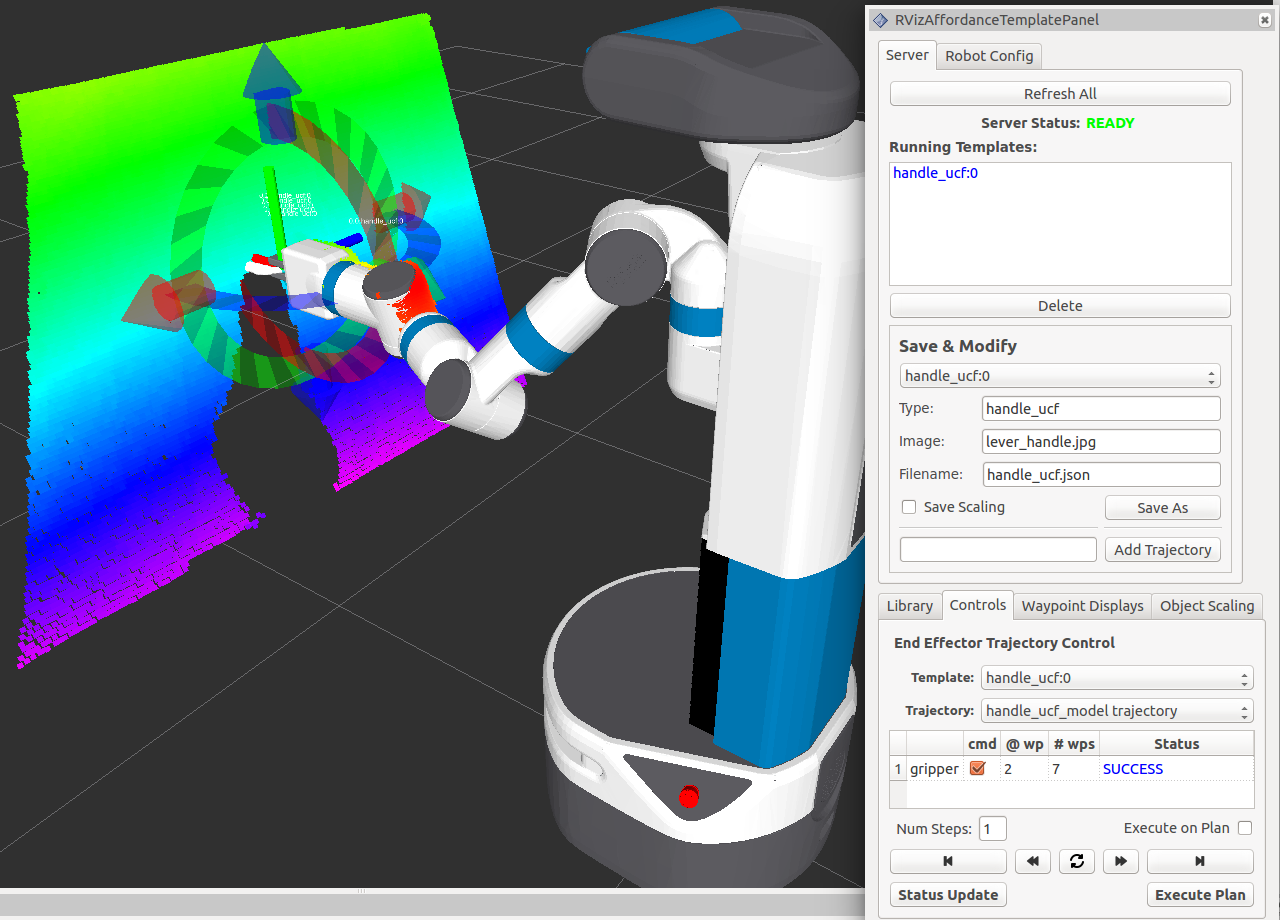}
        \caption{Executing articulation}
        \label{fig:affordance3}
    \end{subfigure}%
    \hspace{0.05cm}
    \caption{Robot performing the task on door handle.}
    \label{fig:affordance}
\end{figure*}

At this point the necessary points for calculating the transformation and articulation for the frames are collected from the user.
Opening a door using a door handle is represented in four frames, moving the gripper to the grasp location, closing the gripper, moving the gripper down toward the projection of the door handle, push or pull the door to open it.

To calculate the frames we need to move the origin and its axes to the object's one. The door handle origin is considered the $p_ah$ and the Y axis lies on $L_r$ and the lever part is along with the X axis. Using this information we calculate the transformation matrix and apply it to the calculated values. 

\subsubsection{Drawer}
To open a drawer, the robot must know where is the handle's grasp point and to which extent the drawer can move out. To provide this data to the robot, the user will create the affordance using our system. In the first step, same as the door handle affordance extraction, the user segments the points in the point cloud by identifying the drawer's surface plane ($P_d$). The grasp point's location ($p_g$) is then drawn by the user; this points projection on $P_d$ is used to create a 3D perpendicular helper and extract the drawer's affordance waypoint for the opened drawer ($p_e$), 

The extracted affordance waypoint is used to create the affordance template for the drawer.

\subsection{Performing The Task on The Robot}


With the poses and geometry from the sketching system prepared, a new template for the robot to perform tasks has to be created and compiled. By executing an ROS command, the JSON file corresponding to the poses is automatically generated using the method shown in previous section. It is then move it to the affordance template package together with the geometry file, creating a new template that is going to finish the task.

With the robot standing in front of the door as its camera is looking at the door handle, as the user open RViz, the newly created door handle task template would be found in the library and is ready to be loaded now as the Figure \ref{fig:affordance1} shows. 

The loaded template has interactive markers to conduct translations, rotations and scaling. By manually applying these transformations on the affordance template, the object in the point cloud scene is to be fitted with the geometry of the affordance template as is illustrated in the Figure \ref{fig:affordance2}. 

After the affordance template's geometry is fully registered with the scene's point cloud, the status window will show "Plan -\textless id[0]" to the user. At this point, the user can start executing the waypoints by clicking on the button. Waypoints are executed one by one by default to keep the execution process safer. It is also possible to change the number of waypoints being executed before user's interruption to any given number. The execution window is shown in the Figure \ref{fig:affordance3}. 

\begin{figure}[h]
  \centering
    \begin{subfigure}[b]{0.48\linewidth}
        \includegraphics[width=\textwidth]{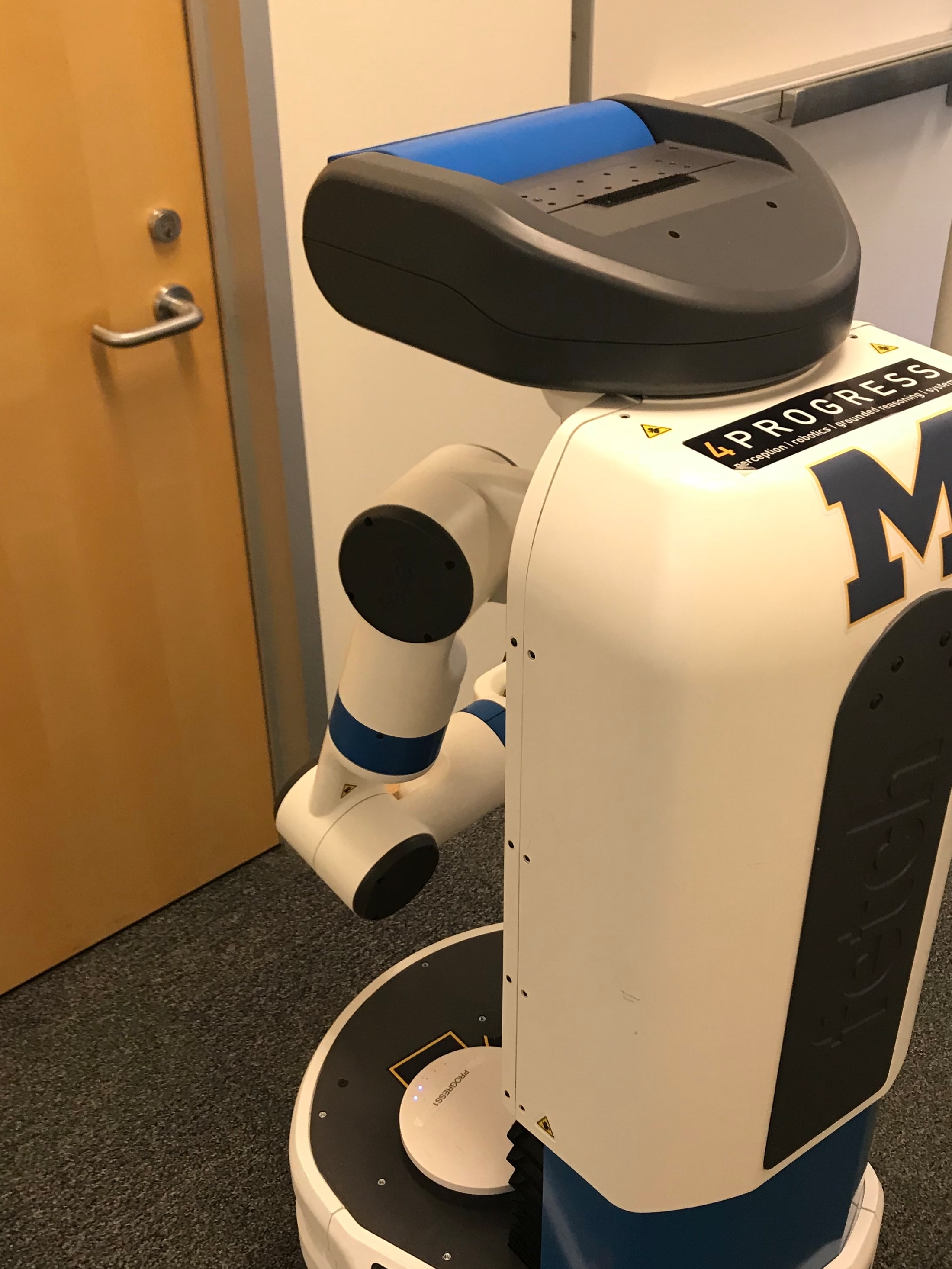}
        \caption{Prepare for the task}
        \label{fig:result1}
    \end{subfigure}%
    \hspace{0.05cm}
    \begin{subfigure}[b]{0.48\linewidth}
        \includegraphics[width=\textwidth]{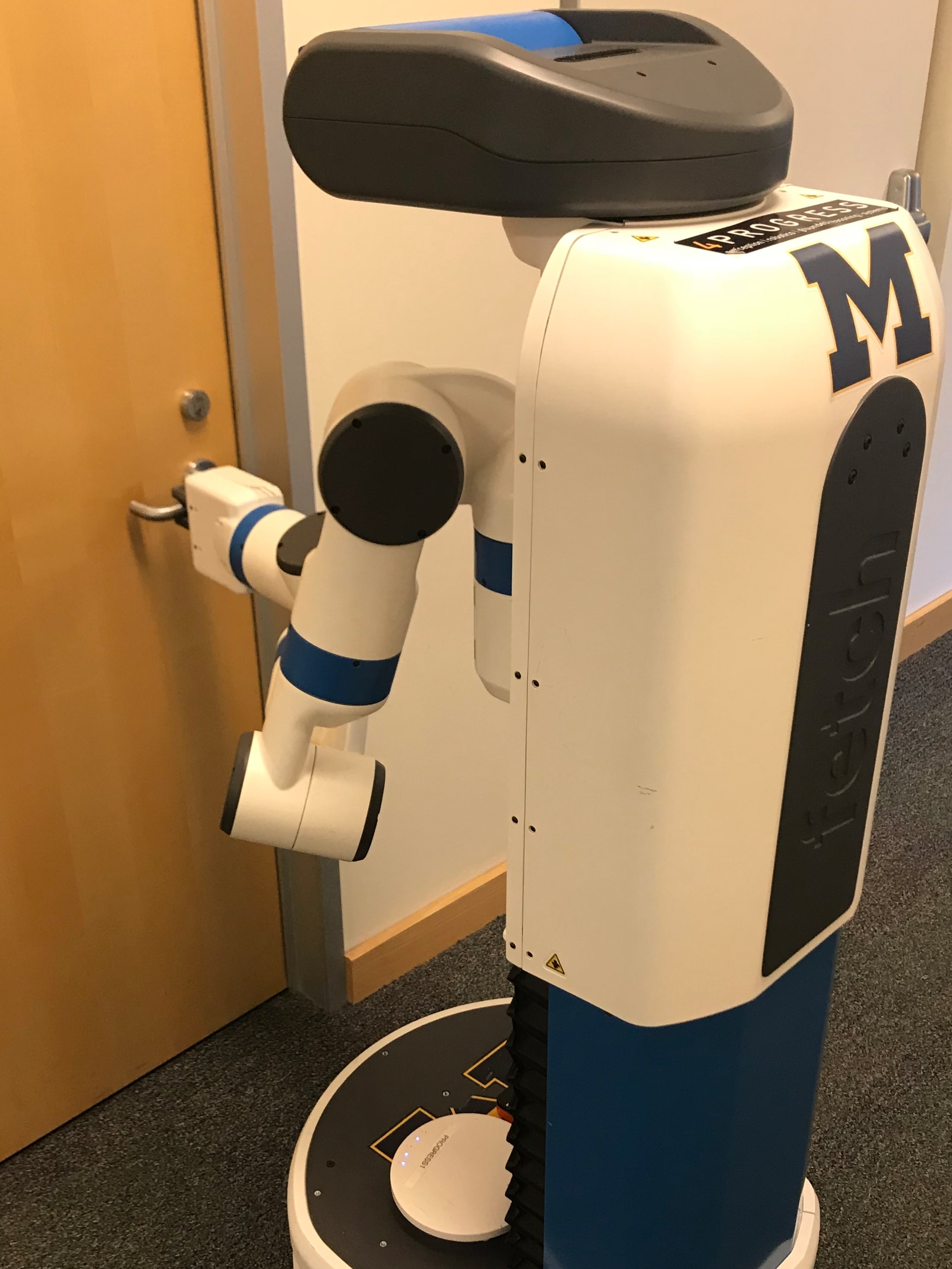}
        \caption{Performing task}
        \label{fig:result2}
    \end{subfigure}%
    \hspace{0.05cm}
    \caption{Robot performing the task on door handle.}
    \label{fig:result}
    
\end{figure}

\begin{figure*}
  \centering
    \begin{subfigure}[b]{0.32\linewidth}
        \includegraphics[width=\textwidth]{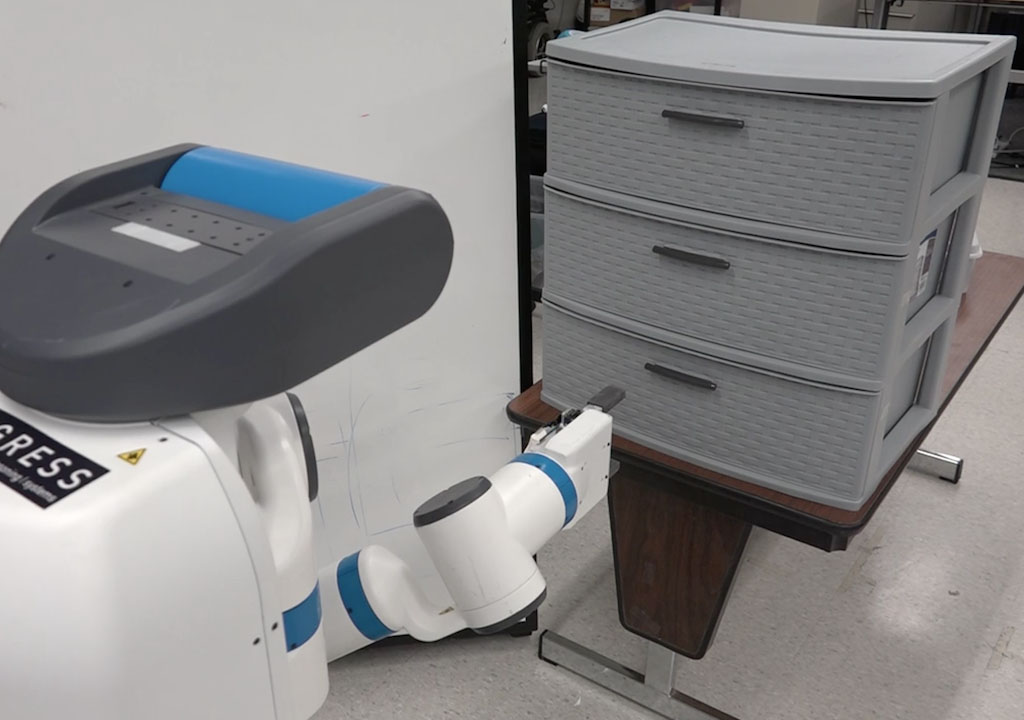}
        \caption{Prepare for opening the drawer}
        \label{fig:drawer1}
    \end{subfigure}%
    \hspace{0.05cm}
    \begin{subfigure}[b]{0.32\linewidth}
        \includegraphics[width=\textwidth]{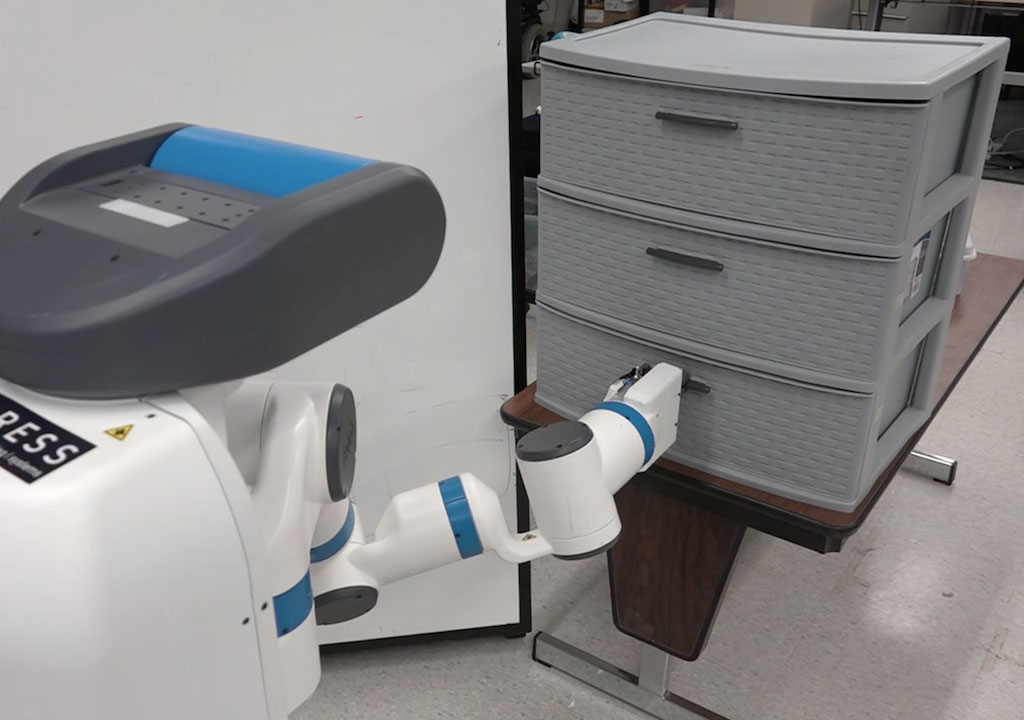}
        \caption{Grabbing the drawer's handle}
        \label{fig:drawer2}
    \end{subfigure}%
    \hspace{0.05cm}
    \begin{subfigure}[b]{0.32\linewidth}
        \includegraphics[width=\textwidth]{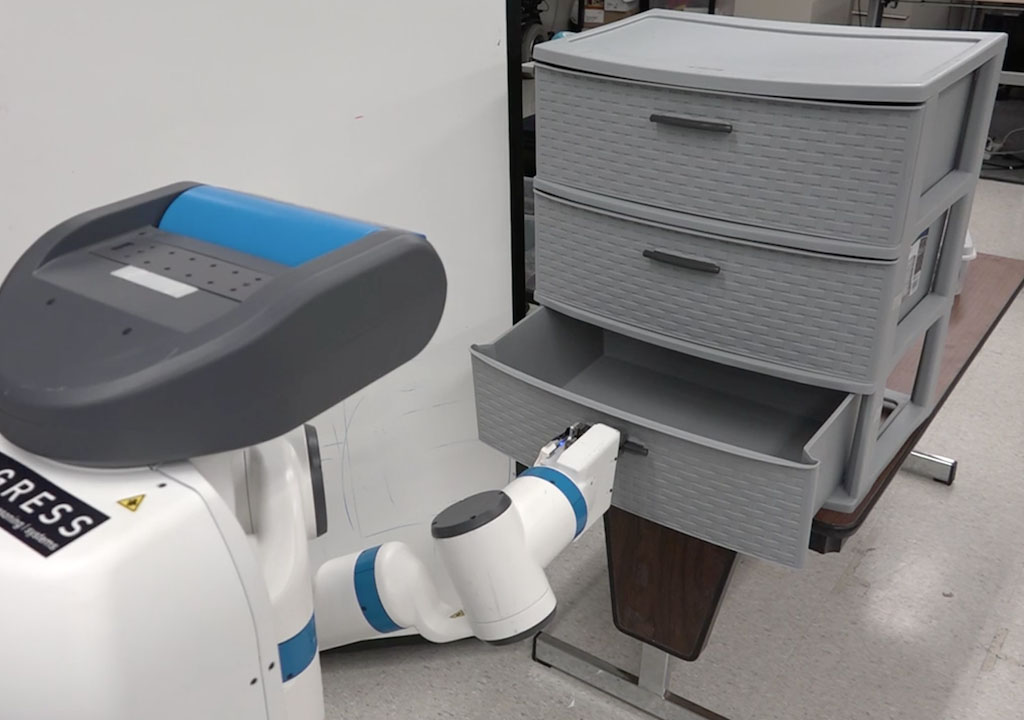}
        \caption{Opening the drawer}
        \label{fig:drawer3}
    \end{subfigure}%
    \hspace{0.05cm}
    \caption{Robot opening a drawer using the provided affordance template}
    \label{fig:drawer}
\end{figure*}

There are also situations that the status window shows "No Plan", which indicates no motion plan can be found by the robot. In these situations the robot should move to a new location and redo the manually fitting geometry process. Figure \ref{fig:result} shows the execution of door handle rotation task and Figure \ref{fig:drawer} shows the robot opening the drawer using the provided data.

\subsection{Result Analysis}
To analyze our system, we employed the sketch system and performed a task of opening a drawer and another task of rotating a door handle. We extracted the geometry and affordance templates ten times for each of them and used the extracted data to perform the tasks.

Using all-or-nothing approach \cite{wolin2008shortstraw} we defined a successful task a task that reaches the final goal successfully. In the case of the door handle, the robot has to grab the door handle and rotate it until the latch is released, and in the case of the drawer, the robot needs to grab the handle and pull it to open. If the robot cannot reach the goal it is considered a failure.

The performance of the robot on the ten extracted data sets for the door handle showed that the robot could reach the goal on all ten date sets successfully. Using the ten extracted data sets for the drawer, the robot could perform the task successfully for all ten data sets.

\section{Discussion}
\label{sec:discussion}

Generating affordance templates for objects without writing codes and creating complex tasks can open door to new possibilities, to use this data and employ this system to create easier robot interactions. Our system can be used by novice users in comparison to the related works. The shared autonomy approach where user aids the robot to access the object's affordances even in a cluttered scene and generating the geometry and affordances by simple sketching, gives our system its unique characteristics.

We have successfully extracted the geometry of a door handle and a drawer handle and used the sketched affordances to generate affordance templates for them and performing the specified tasks.


\vspace{1mm}
\noindent
\vspace{1mm}
\noindent
\textbf{Limitations.~} Our sketch-based affordance extraction system is currently limited to affordances that consist of linear movements and rotation around arbitrary axis. The system can currently provide affordances for robots with a single end effector. The other limitation of our system is that the geometry extraction system is limited to shapes that are a form of cuboids or generalized cylinders. This will limit us for generating affordance templates for more complex shapes.

\section{Conclusion and Future Work}
We introduced a sketch-based system for human guided object manipulation based on the point cloud input. The user interface based affordance template generation provides an easy to use system compared to writing an affordance template. The primitive result shows that our system can successfully be used to extract geometries of generalized cylindrical and cuboid shapes, extract their affordances using sketches, and generate affordance templates for them. The extracted data can be used to perform tasks on a robot. We tested our system by sketching the affordance of a door handle and a drawer handle, generating their affordance templates, providing the robot with the affordance templates, and finally, running the tasks on the robot.

In the future we want to focus on transforming the extracted affordances of one object to other objects. Machine learning techniques can help us apply the knowledge about affordances of one object to a similar one. Our system can help us to ask users extract the affordances of different objects, this task can be done quickly and can be scaled, so we can acquire a huge amount of data and it will be useful to relate these data to other objects.

\addtolength{\textheight}{-12cm}   








References are important to the reader; therefore, each citation must be complete and correct. If at all possible, references should be commonly available publications.

\bibliographystyle{IEEEtran}
\bibliography{references,cite}

\end{document}